\documentclass[aps,reprint,graphicx,superscriptaddress]{revtex4-1}
\usepackage{graphicx}
\usepackage{subfigure}
\usepackage{hyperref}
\usepackage{amsmath}
\usepackage{amssymb}
\usepackage{epstopdf}
\usepackage{bbm}
\usepackage{threeparttable}
\usepackage{appendix}
\usepackage{physics}
\usepackage{float}
\usepackage{cleveref}
\usepackage{textcomp}
\usepackage{xcolor}

\begin{document}

\title{Nonlinear differential imaging via vectorial parametric interaction}

\author{Zhuohang Wei}
\affiliation{State Key Laboratory of Precision Spectroscopy, and Hainan Institute, East China Normal University, Shanghai 200062, China}

\author{Kun Huang}
\email{khuang@lps.ecnu.edu.cn}
\affiliation{State Key Laboratory of Precision Spectroscopy, and Hainan Institute, East China Normal University, Shanghai 200062, China}
\affiliation{Chongqing Key Laboratory of Precision Optics, Chongqing Institute of East China Normal
University, Chongqing 401121, China} 
\affiliation{Innovation Center of Extreme Optics, Shanxi University, Taiyuan, Shanxi 030006, China}

\author{Heping Zeng}
\affiliation{State Key Laboratory of Precision Spectroscopy, and Hainan Institute, East China Normal University, Shanghai 200062, China}
\affiliation{Chongqing Key Laboratory of Precision Optics, Chongqing Institute of East China Normal
University, Chongqing 401121, China}

\begin{abstract}
Optical image differentiation is a key operation for edge extraction in imaging and machine vision, yet most existing implementations rely on momentum-domain filtering elements and are typically developed within a scalar-wave framework. Here we demonstrate a nonlinear vectorial mechanism for optical image differentiation based on parametric wave mixing. By solving the full vector wave equation with a nonlinear polarization source term, we show analytically that frequency conversion intrinsically generates cross-polarized field components that correspond to spatial derivatives of the incident field. Exploiting the polarization-selective phase-matching conditions of second-order nonlinear crystals, particularly uniaxial crystals, we propose filter-free imaging schemes that simultaneously perform spatial differentiation and wavelength conversion. This nonlinear vector differentiation platform enables compact, wavelength-agile, and edge-enhanced imaging, offering new opportunities for mid-infrared imaging and all-optical signal processing.
\end{abstract}

\maketitle

\section{Introduction}

Optical image differentiation, which extracts object edges from images, is a fundamental operation in machine vision, label-free biomedical imaging, remote sensing, and materials characterization \cite{Maurer2011LPR}. Compared with electronic approaches, optical edge detection offers substantially higher processing speed and lower power consumption \cite{Hu2024NP}, making it well suited for integration into all-optical computing platforms or as a front-end processor in advanced imaging architectures \cite{Wesemann2021APR}. Motivated by these advantages, a variety of optical implementations for image differentiation have been developed, each based on distinct physical principles and associated practical trade-offs. Conventional optical differentiation schemes are predominantly based on spatial-frequency filtering, most notably using 4f systems with amplitude or phase masks placed in the Fourier plane \cite{Yan2025APL}. More compact planar realizations engineer the angular response of the system, often described by the optical transfer function of multilayer films \cite{Chazot2020NP} or nanophotonics structure \cite{Wesemann2021Light}. While effective, these approaches rely on external filtering elements and are typically formulated within a scalar-wave approximation, thereby overlooking the vectorial nature of light. In addition, they usually require precise fabrication and alignment of the filtering components \cite{Jesacher2005PRL,Bernet2006OE}, and their performance can be sensitive to wavelength or polarization variations \cite{Dias2025ACSP, Zhou2020SciAdv}.

Beyond momentum-domain filtering, a growing class of optical differentiation schemes exploits the vectorial properties of light \cite{He2022Nanophotonic}. A representative example is differential interference contrast (DIC) microscopy \cite{Tamada2017NC}, which employs birefringent prisms and crossed polarizers to convert phase or amplitude gradients into measurable intensity contrast \cite{Wang2023NC}. Related vectorial mechanisms include differentiation based on the spin Hall effect of light, where spin-orbit coupling induces polarization-dependent spatial shifts that can be harnessed for edge extraction at dielectric interfaces or metasurfaces \cite{Tang2024ACSP}. Polarization-selective effects near the Brewster angle have also been shown to suppress low spatial frequencies, enabling edge enhancement in reflection or transmission geometries \cite{Youssefi2016OL}. Among these vectorial approaches, an especially elegant paradigm exploits the intrinsic generation of cross-polarized components in a tightly focused, linearly polarized beam with a nonplanar wavefront \cite{Zhao2022LPR}. Theoretical analyses reveal that the cross-polarized field is proportional to a spatial derivative of the incident field \cite{Aiello2014AmericanJ}, allowing optical differentiation to be realized through simple polarization filtering \cite{Song2022ACSP}. This mechanism is robust and remains effective under broadband or even incoherent illumination \cite{Zhao2025LPR}. Owing to its simplicity and flexibility, it has attracted increasing interest in applications such as precision metrology \cite{He2024PRApp} and quantitative phase reconstruction \cite{Wang2022PRApp}, particularly in scenarios requiring multimodal switching \cite{Liu2022PRL}.

Despite these advances, research on intrinsic vectorial differentiation has thus far been confined to the linear optical regime. Extending this concept into nonlinear optics offers the prospect of significantly expanded functionality and application space \cite{Barh2019AOP}. Parametric processes such as sum-frequency generation can transfer image information across widely separated spectral bands \cite{Dam2012NP}, a capability that is especially attractive for long-wavelength, room-temperature detection, where indirect imaging using mature visible or near-infrared (NIR) sensors becomes possible \cite{Tan2018NanoP}. Previous studies have combined nonlinear imaging with differential filtering through structured pump beams carrying vortex phases \cite{Qiu2018Optica}, and have demonstrated a range of functional imaging modalities, including high-order differentiation \cite{Zhang2025LPR}, high-throughput imaging \cite{Liu2019PRapp}, and reconfigurable nonlinear filtering based on angular selectivity \cite{Wei2024OL}. Collectively, these efforts have established upconversion-based edge detection as a promising route toward high-precision mid-infrared (MIR) imaging \cite{Zeng2023LPR, Wang2021LPR}. Generally, parametric interactions in nonlinear media also introduce intrinsic polarization selectivity governed by the nonlinear susceptibility tensor and phase-matching conditions \cite{Zhu2022Optica, Ashik2019PR}.  These polarization-related characteristics have found applications in studies of biological tissues \cite{Tuer2012BiophyJournal, Aghigh2023BiophyRev}, crystal polarization dynamics \cite{Lourens2025RevSciInstrum}, and micro- and nanostructure fabrication \cite{Angulo2025AdvMaterInterfaces}. However, to the best of our knowledge, intrinsic vectorial differential imaging has been less studied in the context of second-order nonlinear parametric processes. A proper treatment of this problem requires accounting for the vectorial evolution of the optical field during nonlinear interaction. 

In this work, we establish such a framework and propose a nonlinear differential imaging scheme based on vectorial parametric interactions. By solving the vector wave equation with a nonlinear polarization source term, we analytically demonstrate that frequency conversion intrinsically generates cross-polarized field components possessing spatial differentiation characteristics. Building on this mechanism, we propose and analyze practical imaging schemes that exploit the polarization-selective response of nonlinear crystals, particularly uniaxial crystals, to achieve optical differentiation without any additional  Fourier-domain filtering elements while simultaneously converting the image wavelength. Numerical simulations validate the theoretical predictions by revealing the evolution of polarization components and the resulting edge-enhancement performance. This work provides a compact and physically transparent approach to nonlinear differential imaging, with particular relevance to MIR imaging and vectorial field manipulation in nonlinear photonics.

\section{Theory of Vectorial Differentiation}
\subsection{Nonlinear Parametric Interactions}

The propagation of light in a nonlinear medium is governed by Maxwell's equations, in which the second-order nonlinear polarization $\vb{P}^{(2)}$ acts as a source term in the wave equation:
\begin{equation}
\nabla(\nabla \cdot \vb{E})-\nabla^2\vb{E}
+\mu\epsilon\frac{\partial^2 \vb{E}}{\partial t^2}
=-\mu \frac{\partial^2 \vb{P}^{(2)}}{\partial t^2}\ ,
\label{eq01}
\end{equation}
where $\epsilon$ and $\mu$ denote the permittivity and permeability of the medium, respectively. The nonlinear optical response is determined by both the material properties and the characteristics of the interacting optical fields. In this work, we focus on second-order nonlinear interactions involving two input fields.

As a representative example, we consider sum-frequency generation (SFG), in which a signal photon and a pump photon mix to generate an output photon at a higher frequency. The second-order nonlinear polarization associated with SFG, which changes with the temporal and spatial characteristics of the excitation field, can be written as
\begin{equation}
	\begin{aligned}
	{P}_{i}^{(2)}(\vb{r},\omega_u)
	=\epsilon_0 \sum_{j k}
	\chi^{(2)}_{i j k}(\vb{r};\omega_u;\omega_s,\omega_p) \\
	{E}_{s,j}(\vb{r},\omega_s){E}_{p,k}(\vb{r},\omega_p)\ ,	
	\end{aligned}
	\label{eq02}
\end{equation}
where $\chi^{(2)}_{i j k}$ denotes the second-order nonlinear susceptibility tensor, and $\{i,j,k\}$ label the Cartesian components $\{x,y,z\}$. A spatially modulated second-order nonlinear susceptibility tensor can help satisfy momentum conservation via quasi-phase matching. Here, ${E}_s$ and ${E}_p$ represent the signal and pump fields propagating in the nonlinear crystal, respectively, and ${E}_u$ denotes the generated sum-frequency field. Energy conservation requires that the angular frequencies satisfy $\omega_u=\omega_s+\omega_p$.

In principle, the electromagnetic field at an arbitrary spatial position can be obtained by explicitly expanding the gradient and Laplacian operators in Eq. (\ref{eq01}). However, this procedure leads to three coupled vector wave equations containing mixed spatial derivatives, which makes direct analytical or numerical treatment cumbersome, particularly in the nonlinear regime. For completeness, the full expansion of the vector wave equations is provided in Appendix A.

\subsection{Optical Differentiation via Vectorial Effects}

The Hertz vector potential provides an efficient framework for analyzing electromagnetic wave propagation in both linear and nonlinear media. By introducing the Hertz potential as an intermediate variable, the originally coupled vector wave equations can be transformed into a Helmholtz-type equation, while the physical electric field is recovered through well-defined differential relations \cite{Stratton2007Wiley}. This approach enables a transparent description of polarization coupling induced by wave propagation.

The Hertz electric potential $\vb{\Pi}$ is defined through the relations $\vb{A}=\mu\epsilon \partial \vb{\Pi}/\partial t$ and $\varphi=-\nabla\cdot\vb{\Pi}$, where $\varphi$ and $\vb{A}$ denote the scalar and vector potentials, respectively. The electromagnetic fields are given by $\vb{E}=-\nabla\varphi-\partial\vb{A}/\partial t$ and $\vb{B}=\nabla\times\vb{A}$. A detailed derivation is provided in Appendix B. Here, we focus on the electric field and consider a monochromatic Hertz potential of the form $\vb{\Pi}=\mathrm{Re}\{\widetilde{\vb{\Pi}}(x,y,z)\exp(-i\omega t)\}$. The corresponding Hertz potential satisfies the inhomogeneous Helmholtz equation
\begin{equation}
\nabla^2\widetilde{\vb{\Pi}}+k^2\widetilde{\vb{\Pi}}
=-\frac{\widetilde{\vb{P}}^{(2)}}{\epsilon} \ ,
\label{eq03}
\end{equation}
where $k=n\omega/c$.

The Hertz potential can be expressed as the product of a constant polarization vector $\vb{p}$ and a scalar function $\widetilde{\psi}$ that satisfies the corresponding wave equation under different excitation conditions
\begin{equation}
\widetilde{\vb{\Pi}}=\widetilde{\psi}\vb{p}\ .
\label{eq04}
\end{equation}
In practical imaging systems, the transverse polarization vector is of central interest. Under the paraxial approximation, the longitudinal component along the optical axis is significantly smaller than the transverse component and is therefore justifiably neglected. Linear polarization beam can be represented by a transverse unit vector, $\vb{p}=p_x\vb{e}_x+p_y\vb{e}_y$, where $p_x^2+p_y^2=\cos^2\alpha+\sin^2\alpha=1$. $\alpha$ is the rotation angle of polarization. We can define $\widetilde{\Pi}_x=\widetilde{\psi}p_x$ and $\widetilde{\Pi}_y=\widetilde{\psi}p_y$ in the following text. Once $\widetilde{\vb{\Pi}}$ is determined, the electric field components can be obtained directly. The transverse electric field components are given by
\begin{equation}
	\begin{bmatrix}
		\widetilde{E}_x\\
		\widetilde{E}_y\\
	\end{bmatrix}=
	\mathcal{P}
	\begin{bmatrix}
		\widetilde{\Pi}_x\\
		\widetilde{\Pi}_y\\
	\end{bmatrix}
	+ k^2
	\begin{bmatrix}
		\widetilde{\Pi}_x\\
		\widetilde{\Pi}_y\\
	\end{bmatrix}\ ,
	\label{eq05}
\end{equation}
where the projection operator is defined as
\begin{equation}
	\mathcal{P}=
	\begin{bmatrix}
		\partial_{x}^2 & \partial_{x}\partial_{y} \\
		\partial_{x}\partial_{y} & \partial_{y}^2
	\end{bmatrix}\ .
	\label{eq06}
\end{equation}

To elucidate the vectorial origin of optical differentiation, we consider an incident field polarized along the $x$ direction, $\alpha=0^{\circ},\ p_x=1, \ p_y=0$. In the spatial-frequency domain, the electric field can be written as $\hat{\vb{E}}=\hat{E}_x\vb{e}_x+\hat{E}_y\vb{e}_y=\left[(k^2-k_x^2)\vb{e}_x-k_xk_y\vb{e}_y\right]\hat{\psi}$, revealing an intrinsic coupling between orthogonal polarization components. And the Fourier-transform identity $\mathcal{F}(\partial/\partial x)=ik_x$ has been used. In the plane wave regime, $k_x=k_y=0$, the term $\hat{E}_x\vb{e}_x=k^2\vb{e}_x$ dominates and the field remains predominantly co-polarized with $\vb{p}=\vb{e}_x$. As diffraction effect becomes pronounced (for example, under tight focusing), the transverse wave-vector components $k_x$ and $k_y$ increase, leading to the generation of a cross-polarized field component. Taking into account the actual experimental conditions and the simplicity of the mathematical treatment, we consider the paraxial region, where the angular spectrum of the optical field is narrowly confined around the propagation direction, $k_x,k_y\ll k$ but not a zero value. Thus, the field varies slowly along the transverse directions. For a quantitative analysis, we discuss the off-axis angle of $\theta<10^\circ$ in this work. Thus $k_x/k_0 \approx \sin\theta<0.17$, which can cause a change in the total polarization state.

Considering a nonplanar wavefront, the electric field $\widetilde{\vb{E}}$ naturally decomposes into a co-polarized component $\widetilde{E}_{\parallel}$ and a cross-polarized component $\widetilde{E}_{\perp}$. Their spatial-frequency representations are given by
\begin{equation}
\hat{\vb{E}}_{\parallel}=(k^2-k_x^2)\hat{\psi}\vb{e}_x\ ,
\qquad \hat{\vb{E}}_{\perp}=-k_xk_y\hat{\psi}\vb{e}_y \ .
\label{eq07}
\end{equation}
Performing an inverse Fourier transform, one finds that the cross-polarized component is directly related to the spatial derivative of the co-polarized field
\begin{equation}
\widetilde{E}_{\perp}
\propto \frac{\partial^2\widetilde{E}_{\parallel}}
{\partial x\,\partial y} \ .
\label{eq08}
\end{equation}
This result shows that, during propagation, a linearly polarized optical field with a nonplanar wavefront inherently generates an orthogonally polarized component that encodes spatial differentiation of the original field. Although some studies can reached the same conclusions \cite{Aiello2014AmericanJ}, our expression is derived directly from the relationship between Hertz potential and the light field, offering a more straightforward and efficient approach to address imaging differential issues. Additional mathematical details of the vectorial differential effect are provided in Appendix B.

\begin{figure*}[t!]
	\centering
	\includegraphics[width=0.75\textwidth]{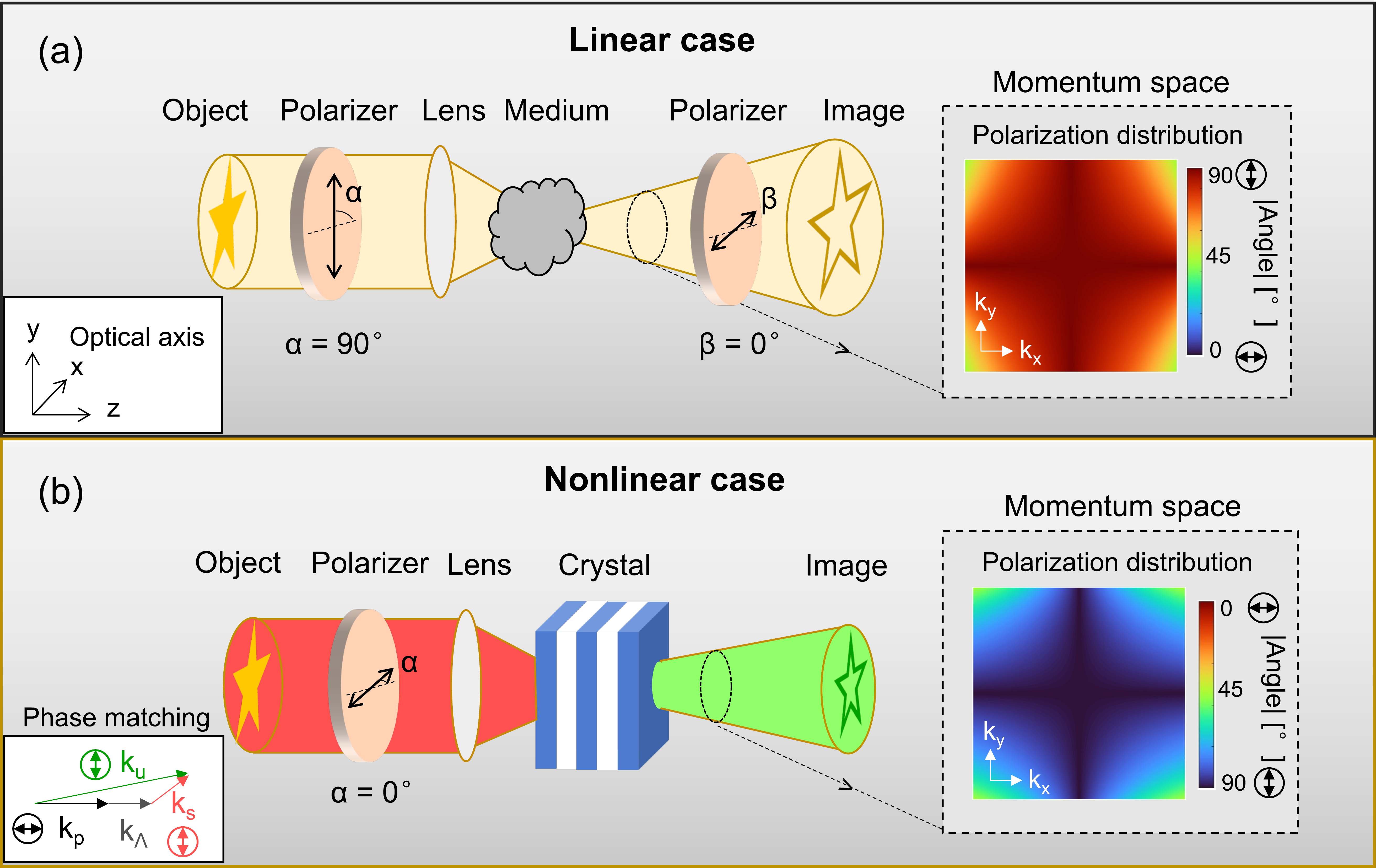}
	\caption{Schematic illustration of vectorial image differentiation in the linear (a) and nonlinear (b) regimes. (a) In the linear configuration, light carrying object information is incident through a polarizer oriented at an angle $\alpha=90^\circ$ relative to the $x$ axis, preserving the spatial-frequency spectrum of the electric field along the incident polarization. Upon focusing, vectorial diffraction induces polarization rotation of spatial-frequency components. The corresponding distribution of polarization rotation in momentum space (shown on the right, without object information) indicates that components orthogonal to the incident polarization are depleted of low spatial frequencies. By introducing a second polarizer at angle $\beta=0^\circ$, low-frequency components are suppressed, enabling edge-enhanced imaging through polarization-based high-pass filtering. (b) In the nonlinear configuration, a uniaxial crystal acts as an intrinsic polarization-selective element that simultaneously enables wavelength conversion. A Type-\MakeUppercase{\romannumeral2} (oeo) phase-matching scheme in a crystal with its optical axis along $x$ axis is considered, with the phase-matching diagram shown at the lower left. When the input polarizer is oriented orthogonally to the phase-matching direction at $\alpha=0^\circ$, only the cross-polarized components carrying edge information of the MIR field are efficiently upconverted to the visible or NIR band. Although the upconverted field also exhibits vectorial polarization mixing, its dominant $y$-polarized component, retains the infrared edge information without requiring additional polarization optics.}
	\label{fig1}
\end{figure*}

\section{Schemes for Differential Imaging}

The crystal structure governs light propagation through both the linear permittivity tensor $\epsilon^{(1)}$ and the nonlinear polarization $\vb{P}^{(2)}$. In the principal coordinate system of a crystal, the relative permittivity tensor takes the diagonal form \cite{Robert2008Academic}
\begin{equation}
	\epsilon^{(1)}  = 
	\begin{bmatrix}
		\epsilon_{xx} & 0 & 0 \\ 
		0& \epsilon_{yy} & 0 \\ 
		0 & 0 & \epsilon_{zz} 
	\end{bmatrix} \ ,
	\label{eq09}
\end{equation}
which provides a distinction of the beam propagation characteristics under different polarization orientation. In practical nonlinear imaging systems, commonly used nonlinear media are predominantly uniaxial crystals, such as LiNbO$_3$ (LN), $\beta$-BaB$_2$O$_4$ (BBO) \cite{Barh2019AOP}. The refractive index depends on the polarization state of the optical field and its orientation relative to the crystal optical axis. In the case that optical axis taken along $x$, which results in $\epsilon_{xx}=n_e^2$ and $\epsilon_{yy}=\epsilon_{zz}=n_o^2$. Such a uniaxial crystal provides a convenient platform to exploit polarization-selective nonlinear interactions.

Beyond linear propagation, a key distinction between media lies in the structure of the second-order nonlinear susceptibility tensor. Efficient frequency conversion is therefore restricted to selected polarization configurations, such as Type-0, Type-\MakeUppercase{\romannumeral1} and Type-\MakeUppercase{\romannumeral2} \cite{Sutherland2003New York}. The phase-matching polarization conditions, causing the nonlinear interaction to act as an intrinsic polarization-selective element. As a representative example, the nonlinear susceptibility tensor of LN crystal (belongs to 3m class) can be written as \cite{Sutherland2003New York}
\begin{equation}
	\begin{bmatrix}
		0 & 0 & 0 & 0 & d_{15} & -d_{22} \\
		-d_{22} & d_{22} & 0 & d_{15} & 0 & 0 \\
		d_{31} & d_{31} & d_{33} & 0 & 0 & 0
	\end{bmatrix}\ ,
	\label{eq10}
\end{equation}
where different tensor elements correspond to distinct polarization combinations. In the following analysis, we focus on the dominant element $d_{31}$ (i.e., $\chi^{(2)}_{oeo}$), which enables Type-\MakeUppercase{\romannumeral2} phase matching with nonlinear polarization
$P^{(2)}=\tfrac{1}{2}\chi^{(2)}_{oeo}E_{s}E_{p}$. Light polarized along the optical axis propagates as an extraordinary wave with refractive index $n_e$. In practical implementations such as periodically poled lithium niobate (PPLN), only a single phase-matching configuration is typically optimized for a given operating condition \cite{Zhu2022Optica}, further reinforcing the intrinsic polarization selectivity of the nonlinear process.

The conceptual framework for differential imaging is illustrated in Fig. \ref{fig1}. In the linear vector-differentiation scheme shown in Fig. \ref{fig1}(a), a plane wavefront illuminates the object, although scattering by the object can alter the polarization state of different spatial-frequency components, a linear polarizer (analyzer) is placed in the path of the scattered field. This ensures that all spatial-frequency components passing through it share a nearly uniform polarization before the Fourier lens. Upon tight focusing in an medium with linear optics property, vectorial diffraction induces polarization rotation that depends on the transverse spatial frequencies. A schematic of the absolute rotation-angle distribution in momentum space (shown on the right) illustrates that low-frequency components experience minimal rotation, while higher spatial frequencies undergo stronger polarization mixing. Such a momentum-dependent rotation of polarization pattern could be experimentally captured directly on the back-focal-plane of a Fourier lens \cite{Song2022ACSP}. By introducing a second polarizer oriented orthogonally to the incident polarization, the low-frequency components are selectively suppressed, resulting in a high-pass filtering effect and edge-enhanced imaging.

In the nonlinear configuration shown in Fig. \ref{fig1}(b), a PPLN crystal simultaneously performs wavelength conversion and polarization-selective filtering. Although birefringence introduces ordinary and extraordinary eigenmodes, the underlying vectorial differentiation mechanism remains valid. Crucially, no second polarizer is required in this scheme. The extinction ratio of polarization devices based on parametric processes can surpass that of conventional Glan prisms. Previous studies have reported that a 3-mm-long nonlinear crystal, when employed as a polarization device, can achieve an extinction ratio on the order of 10$^{-7}$, and this value decreases with the crystal length \cite{Saltiel1987APB}. When the input polarizer is set to $\alpha=0^{\circ}$, the incident MIR field generates cross-polarized components that encode spatial edge information during propagation. Aligning the crystal optical axis along the $x$ direction and establishes the Type-\MakeUppercase{\romannumeral2} (oeo) phase matching, only these cross-polarized components are efficiently upconverted, so that the dominant contribution to the visible or NIR output field originates from the edge information of the MIR image.
 
\begin{figure*}[t!]
	\centering
	\includegraphics[width=0.75\textwidth]{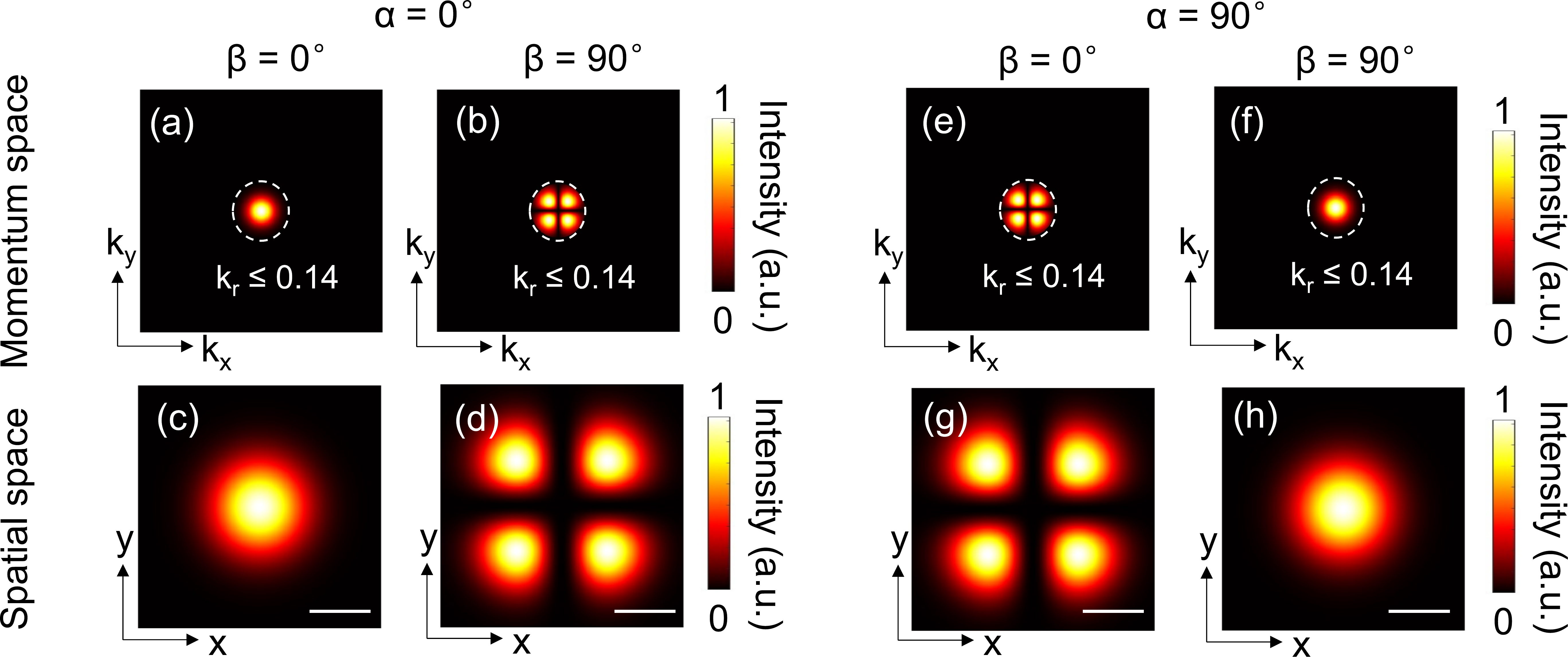}
	\caption{Momentum- and real-space distributions of a Gaussian beam under linear conditions for different combinations of the input polarizer angle $\alpha$ and analyzer angle $\beta$. For $\alpha=0^\circ$ (corresponding to $x$-polarized input), the spatial-frequency distribution at $z$=1 mm with a beam waist $w_0$=20 $\mu$m is shown in (a) for $\beta=0^\circ$, yielding a bright-field response; the corresponding real-space intensity distribution is shown in (c). When the analyzer is rotated to $\beta=90^\circ$, the low spatial-frequency components are suppressed, resulting in the momentum- and real-space patterns shown in (b) and (d), respectively. The corresponding results for $\alpha=90^\circ$ are presented in (e-h), demonstrating the symmetry of the system. Depending on the relative angle $|\alpha-\beta|$, the incident Gaussian beam either retains its bright-field profile or is transformed into a quadrupole pattern, characteristic of spatial differentiation in which the information along both the $x$ and $y$ directions is removed. The scale bar is 50 $\mu$m.}
	\label{fig2}
\end{figure*}
In the following simulations, we select a signal wavelength of 3070 nm (within the molecular fingerprint region) and convert it to 771 nm (for silicon-detector compatibility) as an exemplary pair, which is based on our experimental conditions. This mechanism is wavelength-agnostic and applicable to other spectral ranges under appropriate phase-matching conditions.

\section{Numerical Simulations}
\subsection{Linear Case}
 We first examine the vectorial propagation effect of a MIR beam in the linear regime. In the absence of nonlinear polarization, light propagation is governed by the homogeneous wave equation. Assuming a time-harmonic field with temporal dependence $e^{-i\omega t}$, the wave equation reduces to the Helmholtz equation
\begin{equation}
\nabla^2\widetilde{\vb{\Pi}}+k^2\widetilde{\vb{\Pi}}=0 \ ,
\label{eq11}
\end{equation}
where $\vb{\Pi}$ denotes the Hertz potential introduced in the previous section.

To construct an explicit Hertz potential in Eq. (\ref{eq04}), and calculate the specific light field, we need to specify the scalar function $\psi$ and constant polarization vector $\vb{p}$. We assume a monochromatic and linear polarization Gaussian spatial distribution field, the associated scalar function $\psi$ with the Hertz potential is written as
\begin{equation}
	\psi\left(r,t\right)=E_0\frac{e^{ik_0(z-ct)}}{iQ(z)}\exp\left[-\frac{x^2+y^2}{iw_0^2Q(z)}\right]\ ,
	\label{eq12}
\end{equation}
where $Q(z)=z/L-i$, and Rayleigh range $L=k_0w_0^2/2$. $w_0$ is the beam waist. This formula, through the introduction of a complex parameter $Q(z)$, compactly encapsulates all physical characteristics of a Gaussian beam, namely the evolution of the beam width, the wavefront curvature and the Gouy phase \cite{Aiello2014AmericanJ}.

We select different combinations of polarization angles, obtain the corresponding constant polarization vectors, and calculate the resulting light field distributions in both the momentum and real-space, using Eqs. (\ref{eq04}--\ref{eq06}) and Eq. (\ref{eq12}).

Figure \ref{fig2} summarizes the momentum and real-space distributions of the Gaussian beam after propagation over $z$=1 mm for different combinations of the input polarizer angle $\alpha$ and analyzer angle $\beta$. The beam waist is $w_0$=20 $\mu$m. For the case  $ANG_{c1}=\left\{\alpha,\beta \right\}=\left\{0^{\circ},0^{\circ} \right\}$, the momentum distribution $\hat{E}_{c1}=(k^2-k_x^2)\hat{\psi}$, shown in Fig. \ref{fig2}(a). Low spatial-frequency components aligned with the incident polarization are transmitted, resulting in a conventional bright-field image. In the simulations, the spatial-frequency bandwidth is limited to $k_r/k_0 \leq 0.14$ to mimic a finite numerical aperture, with NA is 0.14 defined in air.  The corresponding spatial distribution of the light field is $\widetilde{E}_{c1}=\left[k^2-4x^2/Q^2(z)w_0^4+2i/Q(z)w_0^2\right]\widetilde{\psi}$, the intensity distribution $|\widetilde{E}_{c1}|^2$ is shown in Fig. \ref{fig2}(c). The scale bar corresponds to 50 $\mu$m. For the case $ANG_{c2}=\left\{0^{\circ},90^{\circ} \right\}$, the momentum and real-space distribution becomes $\hat{E}_{c2}=-k_xk_y\hat{\psi}$, and $\widetilde{E}_{c2}=-4xy\widetilde{\psi}/Q^2(z)w_0^4$. The low-frequency components are selectively suppressed, as shown in Fig. \ref{fig2}(b). The corresponding real-space pattern in Fig. \ref{fig2}(d) exhibits a four-lobed, quadrupole structure, characteristic of a high-pass spatial filtering process in which both $x$- and $y$-directional low-frequency information is removed. Another results for different $\alpha$ and $\beta$ combinations is summarized in Fig. \ref{fig2}(e-h). An equivalent behavior is observed, reflecting the symmetry of the system. In Fig. \ref{fig2}(e, g), $ANG_{c3}=\left\{90^{\circ},0^{\circ} \right\}$, $\hat{E}_{c3}=-k_xk_y\hat{\psi}$ and $\widetilde{E}_{c3}=-4xy\widetilde{\psi}/Q^2(z)w_0^4$, which is similar to the patterns in Fig. \ref{fig2}(b, d). In Fig. \ref{fig2}(f, h), $ANG_{c4}=\left\{90^{\circ},90^{\circ} \right\}$, $\hat{E}_{c4}= (k^2-k_y^2)\hat{\psi}$ and $\widetilde{E}_{c4}=\left[k^2-4y^2/Q^2(z)w_0^4+2i/Q(z)w_0^2\right]\widetilde{\psi}$.

It is important to note that the observed differential effect is governed solely by the relative rotation angle $|\alpha-\beta|$ between the two polarizers, while the absolute orientation of the input polarization can be chosen arbitrarily. For arbitrary input polarization angle $\alpha$ and analyzer angle $\beta$, the output field in the momentum domain is given in Eq. (\ref{eqB8}), Appendix B.

\begin{figure*}[t!]
	\centering
	\includegraphics[width=0.75\textwidth]{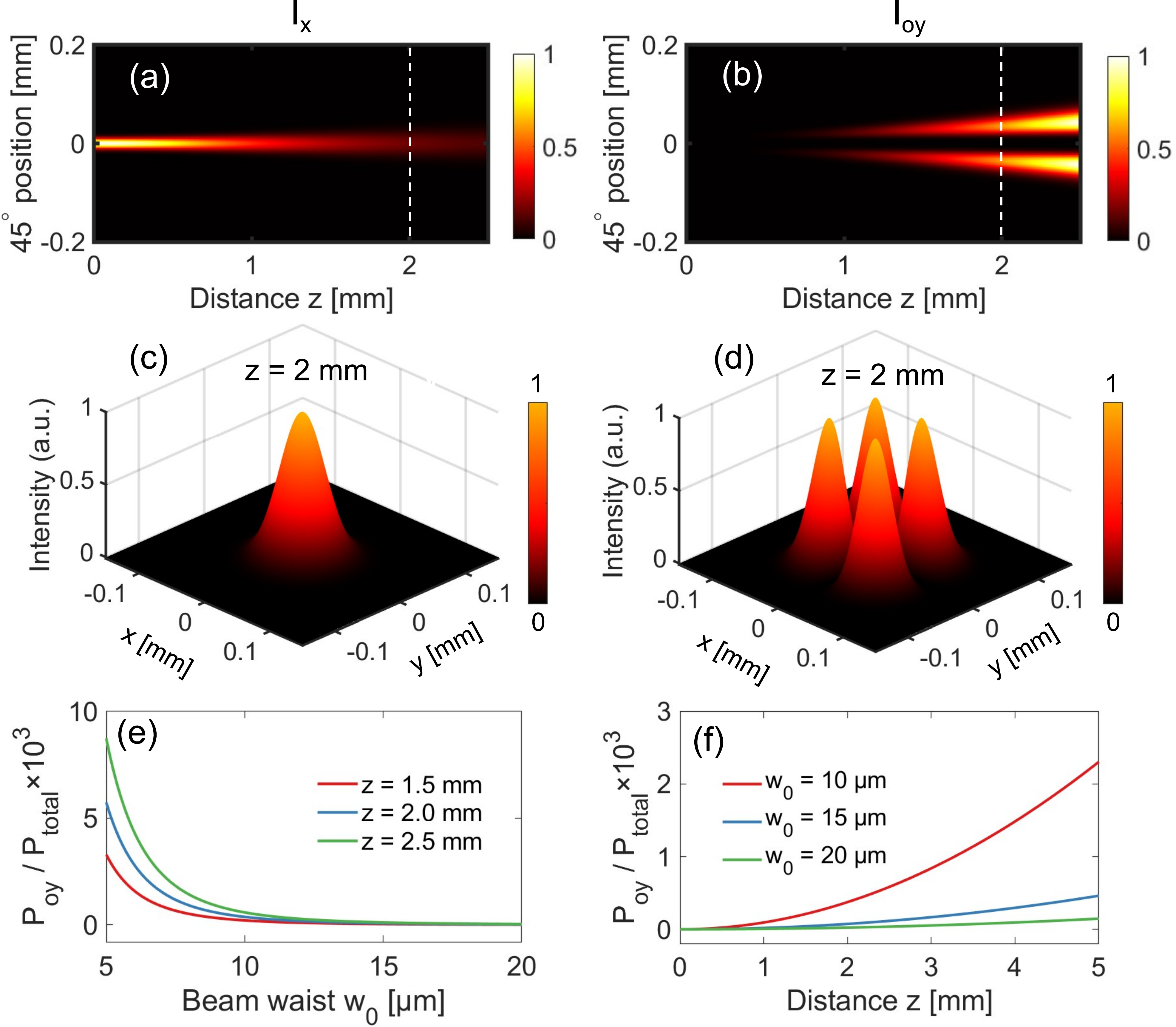}
	\caption{Propagation characteristics of mid-infrared fields ($\lambda$=3070 nm) in a LN crystal. (a,b) Intensity evolution of the co-polarized ($I_{x}$) and cross-polarized ($I_{oy}$) field components along a $45^\circ$ diagonal slice during propagation. (c,d) Corresponding transverse intensity distributions in the $x$-$y$ plane at $z$=2 mm for $I_{x}$ and $I_{oy}$, respectively. (e) Fractional power carried by the cross-polarized component $P_{oy}$ as a function of beam radius for fixed propagation distances of 1.5, 2.0, and 2.5 mm. (f) Fractional power of $P_{oy}$ as a function of propagation distance for fixed beam radii of 10, 15, and 20 $\mu$m. }
	\label{fig3}
\end{figure*}

\subsection{Nonlinear Case}

\subsubsection{Signal Beam Propagation}
We next consider the propagation of the signal beam orthogonal to the optical axis inside a uniaxial crystal, which constitutes the first step of the nonlinear differential imaging process. Here, we choose LN crystal for simulation. The MIR  field propagating inside the crystal can be expressed as a superposition of ordinary and extraordinary modes. In our configuration with the optical axis along $x$, the ordinary wave is polarized perpendicular to the optical axis (i.e., along $y$), corresponding to transverse electric (TE) polarization relative to the $x-z$ plane. The extraordinary wave, polarized within the plane containing the optical axis and the wave vector, corresponds to transverse magnetic (TM) components \cite{Ye2024PRApp}
\begin{equation}
	\vb{E}(\vb{r},z)=\exp(ik_0n_oz)\vb{A}_{o}+\exp(ik_0n_ez)\vb{A}_{e} \ .
	\label{eq13}
\end{equation}
In paraxial regime, the term $\exp(ik_0n_oz)$ and $\exp(ik_0n_ez)$ are the fast-varying carrier describing the rapid phase oscillation along the propagation direction. $\vb{A}_{o}=A_{ox}\vb{e}_x+A_{oy}\vb{e}_y$ and $\vb{A}_{e}=A_{ex}\vb{e}_x+A_{ey}\vb{e}_y$ denote the slowly varying transverse amplitudes of the ordinary and extraordinary waves.

We employ the well-established theoretical framework developed by Ciattoni et al. \cite{Ciattoni2023JOSAA} to analyze signal beam propagation inside the optical  crystal orthogonal to the optical axis, and combine this framework with the specific nonlinear conversion process to further calculate and discuss the propagation characteristics of the signal and upconverted fields. These two modes propagate independently in a uniaxial crystal and are related to the input field through

\begin{equation}
	\begin{bmatrix}
		A_{ox}\\
		A_{oy}
	\end{bmatrix} 
	=\int d^2\vb{k}_\perp
	\mathcal{P}_o
	\begin{bmatrix}
		\hat{E}_x \\
		\hat{E}_y
	\end{bmatrix}
	\exp\left(i\vb{k}_\perp\cdot\vb{r}_\perp\right)\exp(ik_{oz}z)\ ,
	\label{eq14}
\end{equation}
and
\begin{equation}
	\begin{bmatrix}
		A_{ex}\\
		A_{ey}
	\end{bmatrix} 
	=\int d^2\vb{k}_\perp
	\mathcal{P}_e
	\begin{bmatrix}
		\hat{E}_x \\
		\hat{E}_y
	\end{bmatrix}
	\exp\left(i\vb{k}_\perp\cdot\vb{r}_\perp\right) \exp(ik_{ez}z) \ ,
	\label{eq15}
\end{equation}
where $\hat{\vb{E}}=[\hat{E}_x\ \hat{E}_y]^\mathrm{T}$ denotes the two-dimensional Fourier transform of the input transverse field at $z=0$. $\vb{r}_{\perp}=x\vb{e}_x+y\vb{e}_y$, and $\vb{k}_{\perp}=k_x\vb{e}_x+k_y\vb{e}_y$. 
$k_{oz}=(k^2_0n_o^2-k^2_\perp)^{1/2}$, $k_{ez}=(k^2_0n_e^2-n_e^2k_x^2/n_o^2-k^2_y)^{1/2}$.
The Hertz potential derivation reveals that the generation of a cross‑polarized component is an intrinsic consequence of the vectorial nature of light. When this field propagates in a uniaxial crystal, the same differential relationship persists, albeit with modified coefficients that account for birefringence \cite{Ciattoni2023JOSAA, Ciattoni2002JOSAA}. The axial propagation situation is more similar to that derived via the Hertz potential method in an isotropic medium; the analysis can follow the framework in Ref. \cite{Ciattoni2002JOSAA}. In the present orthogonal-to-axis configuration, the form of the projection matrix is given as follows:
\begin{equation}
	\begin{split}
		\mathcal{P}_o&=\frac{1}{k_0^2n_o^2-k_x^2}
		\begin{bmatrix} 
			0 && 0 \\
			k_xk_y && k_0^2n_o^2-k_x^2
		\end{bmatrix}\ ,\\
		\mathcal{P}_e&=\frac{1}{k_0^2n_o^2-k_x^2}
		\begin{bmatrix} 
			k_0^2n_o^2-k_x^2 && 0 \\
			-k_xk_y && 0
		\end{bmatrix}\ .
	\end{split} 
	\label{eq16}		
\end{equation}
An $x$-polarized input, $\hat{\vb{E}}=[\hat{E}_x\ 0]^\mathrm{T}$, can simultaneously excites the ordinary and the extraordinary branch. The excited extraordinary beam with $\mathcal{P}_e\hat{\vb{E}}$ leads to a co-polarization term $(k_0^2n_o^2-k_x^2)\hat{E}_x\vb{e}_x$ with bright field imaging and cross-polarization generation term $-k_xk_y\hat{E}_x\vb{e}_y$ with a characteristic four-lobe structure; the ordinary branch $\mathcal{P}_o\hat{\vb{E}}$ also contribute to the cross-polarization with $k_xk_y\hat{E}_x\vb{e}_y$, as will be quantitatively shown in the following. In contrast, for a $y$-polarized input, $\hat{\vb{E}}=[0\ \hat{E}_y]^\mathrm{T}$, only the bright-field component $(k_0^2 n_o^2 - k_x^2)\hat{E}_y \vb{e}_y$ is present, so the field retains its bright-field character. The transverse electric field can then be reconstructed through Eqs. (\ref{eq13}--\ref{eq15}) and the input condition. 

As a numerical example, we consider an input beam at $z$=0 with polarization parallel to the optical axis, $x$-polarized, and its spatial distribution take a Gaussian form
\begin{equation}
	{\psi}_0(x,y,0)=E_0\exp(-\frac{x^2+y^2}{w_0^2})\ .
	\label{eq17}
\end{equation}
The incident field $\hat{\vb{E}}=\hat{E}_x\vb{e}_x+\hat{E}_y\vb{e}_y$ is
\begin{equation}
	\hat{E}_x=\hat{\psi_0}\ ,\ 	\hat{E}_y=0\ .
	\label{eq18}
\end{equation}
The co-polarization component inside the crystal can be obtained analytically as
\begin{equation}
	\begin{aligned}
		A_{ox}(x,y,z)&=0\ , \\
		A_{ex}(x,y,z)&=\frac{E_0}{\sqrt{Q_{ex}Q_{ey}}}\exp\Bigl[-\frac{x^2}{w_0^2 Q_{ex}}-\frac{y^2}{w_0^2 Q_{ey}}\Bigr]\ .
	\end{aligned}
	\label{eq19}
\end{equation}
Here, $Q_{ex}(z) = 1 + i2 n_e z/k_0 n_o^2 w_0^2$, $Q_{ey}(z)= 1 + i2 z/k_0 n_e w_0^2$. This expression explicitly shows that the diffractive broadening can be different along the two transverse directions. In the case with weak refractive index anisotropy, the broadening effect is negligible under finite propagation distances. The full $x$ polarization field then reads
\begin{equation}
	\begin{aligned}
	E_x(x,y,z)&=A_{ox}e^{i k_0 n_o z}+A_{ex}e^{i k_0 n_e z}\\
	&=A_{ex}(x,y,z)e^{i k_0 n_e z}\ .
	\end{aligned}
	\label{eq20}
\end{equation}
The power $P=\iint|E|^2\,dxdy$ of $x$‑polarized field reads
\begin{equation}
	P_x(z)= \frac{\pi w_0^2 |E_0|^2}{2|Q_{ex}Q_{ey}|}\ .
	\label{eq21}
\end{equation}
The cross-polarized component can be written as
\begin{equation}
	\begin{aligned}
		A_{oy}(x,y,z)&=\frac{-4xyE_{\gamma}}{k_0^2 n_o^2 w_0^4 Q_{o}^3}\exp(-\frac{x^2+y^2}{w_0^2Q_o})\ ,\\
		A_{ey}(x,y,z)
		&=\frac{4xyE_{\gamma}}{k_0^2 n_o^2 w_0^4 Q_{ex}Q_{ey}E_0}A_{ex}(x,y,z)\ .
	\end{aligned}
	\label{eq22}
\end{equation}
Here, $Q_o(z)=Q_{ox}(z)=Q_{oy}(z)= 1 + i2z/k_0 n_o w_0^2$. $E_{\gamma}$ is determined by the incident-field distribution at the input plane and the conservation of the total energy flux \cite{Ciattoni2003PRE}, describes the coupling between the orthogonal and co-polarization components. The cross-polarization power is $P_y\approx P_0(z(n_o-n_e)/n^2k_0w_0^2)^2$ \cite{Brasselet2009OL, Izdebskaya2009OE}, where $n=(n_o+n_e)/2$ . And the y-polarization field component is
\begin{equation}
	E_y(x,y,z)=A_{oy}e^{i k_0 n_o z}+A_{ey}e^{i k_0 n_e z}\ .
	\label{eq23}
\end{equation}
The prefactor $xy$ directly implies a four-lobe transverse intensity distribution for the cross-polarized field, which intensity vanishes along $x=0$ and $y=0$, and peaks in the four quadrants.

In practice, the cross‑polarized field component $E_y$ with phase term as $e^{ik_0(n_o+n_e)z}$, which leading to strong  oscillation during propagation. Nevertheless, the nonlinear process enables selective extraction of the \(A_{oy}\) component, with the phase‑matching condition configured for the Type-\MakeUppercase{\romannumeral2} (oeo), only \(A_{oy}\) with phase term $e^{ik_0n_oz}$ satisfies the phase‑matching requirement for nonlinear conversion. Through phase selection via nonlinear conversion, the cross-polarized optical field suppresses the original phase oscillation, leading to a significantly reduced variation of intensity along the propagation direction. The power of cross-polarized ordinary beam is
\begin{equation}
	P_{oy}(z)=\iint |A_{oy}(x,y,z)|^2 dxdy =\frac{\pi w_0^2 |E_{\gamma}|^2}{32|Q_o|^4} \ .
	\label{eq24}
\end{equation}
The corresponding numerical results are presented in Fig. \ref{fig3}. The temperature is maintained at $24.5^\circ$C to calculate the refractive index of the materials, corresponding to the experimental conditions at room temperature. Figures \ref{fig3}(a) and \ref{fig3}(b) show the evolution of the co-polarized $I_x=|E_x|^2$and cross-polarized $I_{oy}=|E_{oy}|^2$ field components along the propagation direction, respectively, plotted along a diagonal slice. The transverse intensity distributions at $z$=2 mm are shown in Figs. \ref{fig3}(c) and \ref{fig3}(d). The co-polarized component maintains a bright-field, Gaussian-like profile, whereas the cross-polarized component exhibits a characteristic dark-field, quadrupole structure.

For the chosen beam parameters, the Gaussian beam far-field divergence angle $\theta=\lambda/(\pi w_0)$ is sufficiently small to fall within the angular acceptance bandwidth of the phase-matching condition, as determined by the crystal length \cite{Huang2022NC}. This ensures almost constant refractive index across the beam profile and the divergence angle validates the paraxial approximation. The 10° divergence angle corresponds to beam waist radii of 5.6 $\mu m$ at 3070 nm. Figure \ref{fig3}(e) shows the simulated power fraction of the cross-polarized component as a function of beam radius (5-20 $\mu$m) for propagation distances of 1.5, 2.0, and 2.5 mm. The cross-polarized fraction reaches a maximum value of approximately $0.8\%$ and decreases with increasing beam waist. Figure \ref{fig3}(f) displays the dependence of this fraction on propagation distance (0-5 mm) for fixed beam radius of 10, 15 and 20 $\mu$m, revealing a monotonic increase with distance. Although the cross-polarized component accounts for only a small fraction of the total field energy, its extraction through the nonlinear conversion process is experimentally feasible. High-sensitivity upconversion has been demonstrated in our previous work \cite{Huang2021PR}, making detection above the noise floor of standard visible sensors achievable. In the following subsection, we examine how these vectorial MIR components are transferred through nonlinear upconversion.

\begin{figure*}[t!]
	\centering
	\includegraphics[width=0.75\textwidth]{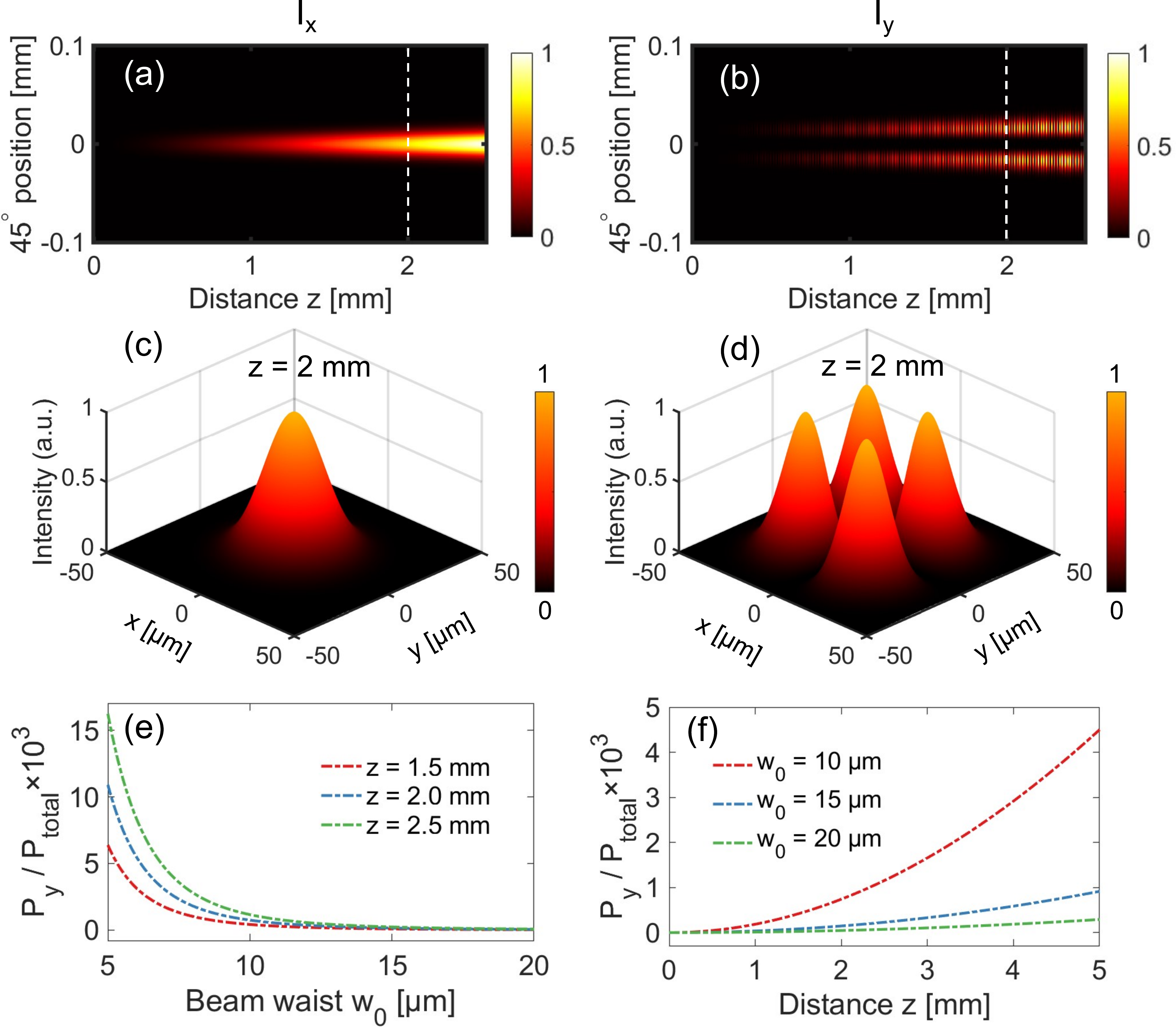}
	\caption{Propagation characteristics of the upconverted field ($\lambda$=771 nm). (a,b) Intensity evolution of the co-polarized ($I_{x}$) and cross-polarized ($I_{y}$) components along the propagation direction, displayed along a $45^\circ$ diagonal slice. (c,d) Corresponding transverse intensity distributions in the $x$-$y$ plane at $z$=2 mm for $I_{x}$ and $I_{y}$, respectively. (e) Fractional power of the cross-polarized component $P_{y}$ as a function of beam radius (5-20~$\mu$m) for fixed propagation distances of 1.5, 2.0, and 2.5 mm. (f) Fractional power of $P_{y}$ as a function of propagation distance for fixed beam radii of 10, 15, and 20 $\mu$m.}
	\label{fig4}
\end{figure*}

\subsubsection{Upconversion Beam Propagation}

In upconversion configuration, the incident signal polarization that carries the MIR edge information is aligned with the selected phase-matching configuration of the crystal.  Our previous analysis, detailed in the paragraph following Eq. (\ref{eq16}), demonstrates that an incident $y$-polarized field does not generate cross-polarization components. Therefore, the edge information encoded in the MIR $y$-polarized field can be directly converted into a visible field with $y$-polarization through Type-\MakeUppercase{\romannumeral2} (oeo) conversion. For completeness, we discuss the other case that upconverted field with $x$-polarized undergoes partial depolarization as a consequence of vectorial diffraction and birefringent propagation.

In numerical treatments of wave equations with nonlinear source terms, the generated field is commonly obtained as a particular solution driven by a prescribed source distribution, which in the present analysis takes the form of a product of two Gaussian functions (signal and pump fields) multiplied by a spatially-dependent coefficient \cite{Boyd1968JAP}. The product of two Gaussian functions remains Gaussian, with their beam waist radii satisfying a certain correlation; consequently, the resulting nonlinear polarization source follows a Gaussian distribution. This source drives the wave equation, yielding an upconversion field inside the crystal that can also be constructed as a Gaussian field with a propagation coefficient \cite{Kleinman1966PR,Boyd1965PR}, as detailed in Appendix C. 

For an initially $x$-polarized upconversion beam, the field at $z$ take the form
\begin{equation}
	\hat{E}_x=C(z)\hat{\psi_0}\ ,
	\qquad
	\hat{E}_y=0\ ,
	\label{eq25}
\end{equation}
where the longitudinal buildup factor $C(z)$ is determined by the nonlinear coupling \cite{Kleinman1966PR}. For mathematical convenience, the beam waist is placed at the crystal entrance face $z=0$. In practice, it can be positioned elsewhere, e.g., focused at the crystal center. Transmission at the crystal-air interface imparts is treated a constant power scaling factor for the small angles, with negligible effect on the spatial differentiation or edge-enhancement characteristics. In fact, the nonlinear process does not change the transverse distribution pattern, but alters the intensity distribution in the direction of propagation. The expression of co-polarization upconversion field can be achieved simply by multiplying the corresponding optical field amplitude in Eq. (\ref{eq20}) and Eq. (\ref{eq21}) by the propagation factor $C(z)$. Under the paraxial and near-phase-matching approximations, and with absorption neglected, the integration over $C(z)$ yields $C(z) \approx \kappa z$ \cite{Boyd1968JAP,Boyd1965PR}. Normalizing out the nonlinear coupling strength $\kappa$ (i.e., dividing by $\kappa$), the cross‑polarization upconversion field is given by
\begin{equation}
	E_y(x,y,z)\approx z(A_{oy}e^{i k_0 n_o z}+A_{ey}e^{i k_0 n_e z})\ .
	\label{eq26}
\end{equation}
Figure \ref{fig4} illustrates the spatial evolution of the two polarization components of the upconverted field arises from a second-order nonlinear process. Higher-order or other type of phase matching nonlinear processes are negligible under this operation conditions due to their significantly lower efficiency and different phase-matching requirements. As shown in Figs. \ref{fig4}(a, b), the co-polarized component retains a Gaussian-like intensity profile during propagation, accompanied by gradual beam expansion. The cross-polarization component oscillates violently with the propagation distance. The corresponding transverse intensity distributions at $z$ = 2 mm are shown in Figs. \ref{fig4}(c,d).

To quantitatively assess the evolution of polarization components, we evaluate the fraction of power carried by the cross-polarized field. The corresponding power is
\begin{equation}
	P_y\approx zP_{oy}\sin^2\left[\frac{k_0(n_e-n_o)z}{2}\right]\ .
	\label{eq27}
\end{equation}
The oscillation cycle is given as $L_p=2\pi/(k_0|n_e-n_o|)$. To illustrate the averaged power envelope evolution, in the simulation we take $\langle P_y \rangle\approx zP_{oy}/2$, thus the fast oscillations are not shown. Fig. \ref{fig4}(e) shows the cross-polarized power fraction as a function of beam radius for fixed propagation distances of 1.5, 2.0, and 2.5 mm. The 10° divergence angle corresponds to beam waist radii of 1.41 $\mu m$ at 771 nm. Fig. \ref{fig4}(f) presents its dependence on propagation distance for fixed beam radii of 10, 15, and 20 $\mu$m. The results indicate that the proportion of the cross-polarized component decreases with increasing beam size, and increases with longer propagation distances inside the crystal. Therefore, the upconversion component carrying edge enhancement information occupies the major part of the power.

These observations demonstrate that the vectorial differential characteristics of light propagation are largely intrinsic and are preserved during nonlinear frequency conversion. The nonlinear source term primarily redistributes energy between polarization components rather than altering the underlying spatial differentiation mechanism, thereby maintaining the feature of edge-enhanced imaging after wavelength conversion.

\begin{figure}[t!]
	\centering
	\includegraphics[width=0.45\textwidth]{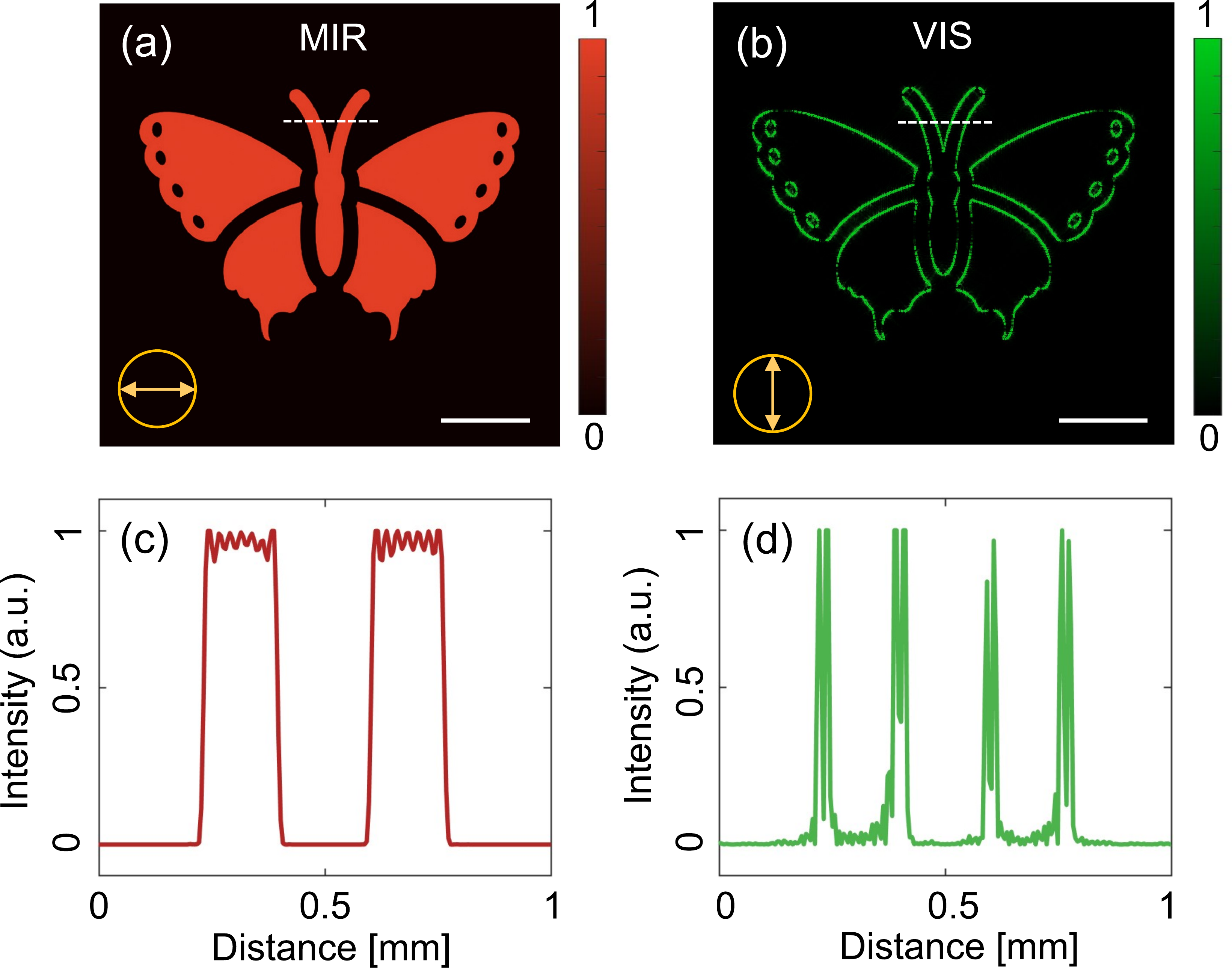}
	\caption{Nonlinear edge-enhanced imaging based on vectorial parametric upconversion. (a) Bright-field image the incident MIR object. (b) Edge-enhanced image of the upconversion field in the visible (VIS) spectral region, due to the selective second-order nonlinear conversion of the cross-polarized components. (c,d) Intensity profiles corresponding to the dashed lines in (a) and (b), respectively. Scale bar: 1 mm.}
	\label{fig5}
\end{figure}

\section{Nonlinear Edge-Enhanced Imaging}

Finally, we evaluate the performance of nonlinear differential imaging in a 4f system through numerical simulations. To reflect realistic experimental conditions, the spatial-frequency bandwidth in our simulations is limited by the numerical aperture of the system, with $k_r/k_0<0.14$. The proposed nonlinear differential operation stems from the intrinsic vectorial filtering imposed by phase matching. It requires only that a linearly polarized beam carrying object information is tightly focused into the crystal, a condition achievable with various imaging geometries (e.g., multi-lens imaging systems) beyond the illustrative 4f configuration in Fig. \ref{fig1}. Consequently, the differentiation imaging technique can be extended to other optical arrangement and does not require any additional Fourier filtering elements.

The nonlinear vector imaging process is modeled by sequentially accounting for the relevant optical interactions. An MIR beam illuminates a test object with a butterfly pattern, passes through an input polarizer, and is subsequently focused into an LN crystal. Figure \ref{fig5} summarizes the resulting MIR and visible images obtained after propagation through the crystal. As shown in Fig. \ref{fig5}(a), the co-polarized ($\vb{e}_x$) MIR component produces a conventional bright-field image, whereas the cross-polarized ($\vb{e}_y$) MIR component selectively retains edge information. Upon propagation through the nonlinear crystal, phase matching acts as an intrinsic polarization-selective filter, such that only the MIR edge components satisfying the phase-matching condition are efficiently upconverted. The MIR edge information undergoes simultaneous wavelength conversion and partial depolarization, resulting upconverted co-polarized ($\vb{e}_y$) visible image in Fig. \ref{fig5}(b), exhibits feature preserved transfer of the MIR edge information into the visible spectral region. Here, the scale bar is 1 mm. The spatial distribution of the upconverted cross-polarized component corresponds to a second-order spatial differentiation of the original pattern. Nevertheless, as established in the preceding sections, the co-polarized component carries the dominant fraction of the total optical energy and therefore provides an effective edge-enhanced image without the need for an additional output polarizer. The corresponding intensity line profile in Fig. \ref{fig5}(c) and \ref{fig5}(d) clearly reveals the enhanced contrast at the object boundaries, confirming the differential nature of vectorial parametric interaction. 

It is worth noting that, the periodically poled nonlinear crystal in 4f system imposes a phase-matching angular acceptance bandwidth that scales inversely with crystal length, thereby limiting the spatial frequency content of the bright-field image. For a numerical aperture of 0.14, the required crystal length is calculated to be around 1 mm. The nonlinear differentiation process preserves the intrinsic spatial resolution and field of view of the imaging system, as these properties are determined by the underlying optical architecture rather than by the differential operation itself. In addition, the cross-polarization-based transfer function is directionally specific \cite{Liu2022NC}, some segments of the edge are missing in Fig. \ref{fig5}(b). Isotropic differentiation may be realized by cascading crystals for parallel multi-mode conversion or considering circularly polarized illumination \cite{Yang2024Optica}.

\section{Conclusion}

In summary, we have theoretically and numerically investigated a nonlinear mechanism for spatial image differentiation based on cross-polarization generation of the input and the up-converted wavelength inside a second-order parametric process. By solving the full vector wave equation, we show that both tightly focused signal fields and parametrically upconverted fields naturally generate orthogonally polarized components during propagation, which are directly linked to spatial derivatives of the co-polarized field. Analysis of polarization evolution and energy redistribution under nonlinear crystal confirms that this vectorial differential effect persists in nonlinear crystals and can be exploited for efficient edge extraction.

Based on this mechanism, we propose a nonlinear vector differentiation scheme that operates without Fourier-domain filtering elements and requires no additional polarization optics beyond a single input polarizer. The scheme relies solely on intrinsic vectorial propagation and polarization-selective phase matching in nonlinear crystals, thereby simplifying the optical architecture while simultaneously enabling wavelength conversion of the edge-enhanced image. In particular, the conversion from the mid-infrared to the visible spectral region allows high-sensitivity detection using mature silicon-based sensors \cite{Tan2018NanoP}. Numerical simulations validate this approach, demonstrating edge extraction with significant contrast enhancement.

The theoretical framework developed here is generic and can be extended to other nonlinear processes, such as difference-frequency generation. It provides a versatile foundation for designing compact, polarization-selective imaging systems in spectral regimes where conventional detectors face intrinsic limitations, and offers new possibilities for vector-field manipulation in nonlinear photonics \cite{Li2017NatRevMat}, structured-light imaging \cite{Singh2024AdvPhoto}, and integrated all-optical signal processing \cite{Wang2024NatCompu}.

\section*{Acknowledgments}
This work was funded by National Key Research and Development Program (2025YFA1411700); National Natural Science Foundation of China (62235019, 62035005, 62505088, 125B2088); Shanghai Pilot Program for Basic Research (TQ20220104); Postdoctoral Fellowship Program (GZC20250545); China Postdoctoral Science Foundation (2024M760918, 2025T180224).

\section*{Conflict of interest}
The authors declare no conflict of interest.

\section*{Data availability statement}

The data that support the findings of this study are available from the corresponding author upon reasonable request.

\section*{Keywords}

Differential imaging; edge enhancement; upconversion imaging; parametric wave mixing; Hertz vector potential

\section*{APPENDIX A: Vector Wave Equation}
\label{app:A}
The standard derivation in this part is excerpted from the cited textbook  for clarity \cite{Stratton2007Wiley}. Maxwell's equations in a nonlinear optical medium can be written as
\begin{equation}
	\begin{aligned}
	\nabla\cdot \vb{D}&= \rho\ ,\\
	\nabla\cdot \vb{B}&= 0\ ,\\
	\nabla\times \vb{E}&=-\frac{\partial \vb{B}}{\partial t}\ , \\
	\nabla\times \vb{H}&=\vb{J}+\frac{\partial \vb{D}}{\partial t}\ ,
	\end{aligned}
	\label{eqA1}
	\tag{A1}
\end{equation}
where $\rho$ and $\vb{J}$ denote the free charge and current densities, respectively. In the absence of free charges and currents ($\rho=0$, $\vb{J}=0$), and neglecting magnetization effects such that $\vb{B}=\mu\vb{H}$ with $\mu=\mu_0\mu_r$ and $\mu_r=1$, the curl of Faraday's law yields
\begin{equation}
	\nabla\times\nabla\times \vb{E}+ \mu\frac{\partial^2 \vb{D}}{\partial t^2}=0\ .
	\label{eqA2}
	\tag{A2}
\end{equation}	

The electric displacement field is given by
\begin{equation}
	\vb{D}=\epsilon \vb{E}+\vb{P}^{\text{(2)}}\ ,
	\label{eqA3}
	\tag{A3} 
\end{equation}
where $\vb{P}^{(2)}$ denotes the second-order nonlinear polarization and $\epsilon=\epsilon_0(1+\chi^{(1)})=\epsilon_0\epsilon_r$,
with $\chi^{(1)}$ the linear electric susceptibility. Using the vector identity $\nabla\times\nabla\times\vb{E} =\nabla(\nabla\cdot\vb{E})-\nabla^2\vb{E}$, Eq. (\ref{eqA2}) can be rewritten as
\begin{equation}
	\nabla(\nabla \cdot \vb{E})-\nabla^2\vb{E}+\mu\epsilon\frac{\partial^2}{\partial t^2}\vb{E}=-\mu \frac{\partial^2 \vb{P}^{(2)}}{\partial t^2}\ .
	\tag{A4}
	\label{eqA4}
\end{equation}

A direct solution of Eq. (\ref{eqA4}) requires explicit expansion of the differential operators. For each Cartesian component $i=x,y,z$, the Laplacian takes the form
\begin{equation}
\nabla^2 E_i
=\frac{\partial^2 E_i}{\partial x^2}
+\frac{\partial^2 E_i}{\partial y^2}
+\frac{\partial^2 E_i}{\partial z^2},
\tag{A5}
\end{equation}
while the gradient of the divergence can be written as
\begin{equation}
\begin{aligned}
\nabla(\nabla\cdot\vb{E})=&
\left(
\frac{\partial^2 E_x}{\partial x^2}
+\frac{\partial^2 E_y}{\partial x\partial y}
+\frac{\partial^2 E_z}{\partial x\partial z}
\right)\vb{e}_x\\
&+
\left(
\frac{\partial^2 E_x}{\partial y\partial x}
+\frac{\partial^2 E_y}{\partial y^2}
+\frac{\partial^2 E_z}{\partial y\partial z}
\right)\vb{e}_y\\
&+
\left(
\frac{\partial^2 E_x}{\partial z\partial x}
+\frac{\partial^2 E_y}{\partial z\partial y}
+\frac{\partial^2 E_z}{\partial z^2}
\right)\vb{e}_z \ .
\end{aligned}
\tag{A6}
\end{equation}

Substituting these expressions into Eq. (\ref{eqA4}) yields three coupled vector wave equations containing mixed second-order derivatives with respect to the longitudinal ($z$) and transverse ($x,y$) coordinates. In the presence of a nonlinear polarization source term, these coupled equations become particularly cumbersome to treat analytically or numerically. This complexity motivates the introduction of alternative formalisms, such as the Hertz vector potential approach employed in the main text, which enables a more transparent and tractable description of vectorial propagation and nonlinear interactions.

\section*{APPENDIX B: Hertz Vector Potential}
\label{app:B}

The analysis of electromagnetic wave propagation can be greatly simplified by introducing auxiliary potential functions \cite{Stratton2007Wiley}. In the conventional formulation, the electromagnetic fields are expressed in terms of the scalar potential $\varphi$ and the vector potential $\vb{A}$ as
\begin{equation}
	\begin{aligned}
	\vb{E}=&-\nabla \varphi-\frac{\partial \vb{A}}{\partial t}\ ,\\
	\vb{B}=&\nabla\times \vb{A}\ .
	\end{aligned}
	\label{eqB1}
	\tag{B1}
\end{equation}

An alternative and particularly convenient formulation is provided by the Hertz vector potential. In this work, we introduce the electric Hertz potential $\vb{\Pi}$, while neglecting the magnetic Hertz potential. The scalar and vector potentials are related to $\vb{\Pi}$ through
\begin{equation}
	\begin{aligned}
	\vb{A}=&\mu\epsilon\frac{\partial \vb{\Pi}}{\partial t}\ ,\\
	\varphi=&-\nabla \cdot \vb{\Pi}\ .
	\end{aligned}
	\label{eqB2}
	\tag{B2}
\end{equation}
Substituting these relations into Eq. (\ref{eqB1}), the electric field can be written directly in terms of the Hertz potential as
\begin{equation}
	\vb{E}=\nabla(\nabla \cdot \vb{\Pi})-\mu\epsilon\frac{\partial^2 \vb{\Pi}}{\partial t^2}\ .
	\tag{B3}
	\label{eqB3}
\end{equation}
In an isotropic and homogeneous medium with no conduction currents or free charges, the Hertz potential satisfies the following homogeneous wave equation
\begin{equation}
	\nabla^2\vb{\Pi}-\mu\epsilon\frac{\partial^2\vb{\Pi}}{\partial t^2}=0\ .
	\tag{B4}
	\label{eqB4}
\end{equation}
For harmonic time dependence, $\vb{\Pi}=\mathrm{Re}\{\widetilde{\vb{\Pi}}(x,y,z)\exp(-i\omega t)\}$, Eq. (\ref{eqB4}) becomes the Helmholtz equation. These detailed derivation process can be found in standard textbooks (see, for example, Reference \cite{Stratton2007Wiley}, page 29-31).

In nonlinear case, the Hertz potential satisfies an inhomogeneous wave equation driven by a second-order nonlinear polarization source. 
\begin{equation}
	\nabla^2\vb{\Pi}-\mu\epsilon\frac{\partial^2\vb{\Pi}}{\partial t^2}=-\frac{\vb{P}^{(2)}}{\epsilon}\ ,
	\tag{B5}
	\label{eqB5}
\end{equation}
where $\vb{P}^{(2)}=\vb{D}-\epsilon\vb{E}$ denotes the polarization density. Formally, a special solution for $\vb{\Pi}$ can be obtain using the Green-function representation. The presence or absence of a nonlinear interaction therefore modifies only the source term of the Hertz potential equation, while preserving the same formal structure for field reconstruction.

The Hertz potential $\widetilde{\vb{\Pi}}=\widetilde{\psi}\vb{p}$ in general polarization vector is $\vb{p}=p_x\vb{e}_x+p_y\vb{e}_y+p_z\vb{e}_z$. In momentum domain, the relation between the electric field and Hertz potential, Eq. (\ref{eqB3}), becomes $\hat{\vb{E}}=-\vb{k}(\vb{k}\cdot\hat{\vb{\Pi}})+k^2\hat{\vb{\Pi}}$, and $\hat{\vb{E}}=\hat{E}_x\vb{e}_x+\hat{E}_y\vb{e}_y+\hat{E}_z\vb{e}_z$, the expression can be rewritten as
\begin{equation}
	\begin{bmatrix}
		\hat{E}_x \\
		\hat{E}_y \\
		\hat{E}_z
	\end{bmatrix}=
	\begin{bmatrix}
		k^2-k_x^2 && -k_xk_y && -k_xk_z\\
		-k_xk_y && k^2-k_y^2 && -k_zk_y\\
		-k_xk_z && -k_yk_z && k^2-k_z^2
	\end{bmatrix}
	\begin{bmatrix}
		p_x\\
		p_y\\
		p_z
	\end{bmatrix}\hat{\psi}\ ,
	\tag{B6}
	\label{eqB6}	
\end{equation}
where $k=n\omega/c$, $n=\sqrt{\epsilon_r\mu_r}$.

A result applicable to any incident polarization angles $\alpha$ and analyzer polarization angles $\beta$ is obtained. Considering only the transverse light field component, the filed behind the incident polarizer with angles $\alpha$, $p_x=\cos\alpha$, $p_y=\sin\alpha$, becomes
\begin{equation}
	\begin{bmatrix}
		\hat{E}_x\\
		\hat{E}_y\\
	\end{bmatrix}=
	\begin{bmatrix}
	(k^2-k_x^2)\cos\alpha-k_xk_y\sin\alpha\\
	-k_xk_y\cos\alpha+(k^2-k_y^2)\sin\alpha\\
	\end{bmatrix}\hat{\psi}\ .
	\tag{B7}
	\label{eqB7}		
\end{equation}
After passing through a analyzer with polarization angles $\beta$, the light field can be calculated as
\begin{equation}
	\begin{split}
		\hat{\vb{E}}_{out}=(\hat{E}_x\cos\beta+\hat{E}_y\sin\beta)\vb{e}_{\beta}=[(k^2-k_x^2)\cos\alpha\cos\beta \\
		+ (k^2-k_y^2)\sin\alpha\sin\beta-k_xk_y\sin(\alpha+\beta)]\hat{\psi}\vb{e}_{\beta}\ .
	\end{split}
	\tag{B8}
	\label{eqB8}
\end{equation}
At small transverse spatial frequencies, $(k^2-k_x^2)\approx (k^2-k_y^2)\approx k^2$, $\hat{\vb{E}}_{out}\approx [k^2\cos(\alpha-\beta)-k_xk_y\sin(\alpha+\beta)]\hat{\psi}\vb{e}_{\beta}$.
When $|\alpha-\beta|=90^{\circ}$, the system exhibits differential imaging operation, and the differential imaging reaches its brightest at $\alpha=$0 or 90$^\circ$.

\section*{APPENDIX C: Approximate Solution for the Upconverted Field}

We consider the propagation of the upconverted field inside the crystal, which is associated with the generated field satisfies an inhomogeneous Helmholtz equation, namely
\begin{equation}
	\nabla^2 \vb{\psi}+k_u^2\vb{\psi}=-\frac{P_{\text{eff}}^{(2)}}{\epsilon}\ .
	\tag{C1}
	\label{eqC1}
\end{equation}
For the sum-frequency component at $\omega_u=\omega_s+\omega_p$,  the corresponding nonlinear polarization, $P_{\text{eff}}^{(2)}=\epsilon_0 d_{\mathrm{eff}} E_s E_p$, can be written as \cite{Boyd1968JAP}
\begin{equation}
	P_u^{(2)}(\mathbf r)=
	2\epsilon_0 d_{\mathrm{eff}}
	A_s(\mathbf r_\perp,z)A_p(\mathbf r_\perp,z)
	\exp[i(\mathbf k_s+\mathbf k_p)\cdot\mathbf r],
	\tag{C2}
	\label{eqC2}
\end{equation}
where $d_{\mathrm{eff}}$ is the effective nonlinear coefficient. $A_s(\vb{r}_\perp)$ and $A_p(\vb{r}_\perp)$ denote the envelopes of the signal and pump fields, respectively. A formal particular solution of Eq. (\ref{eqC1}) can then be written in the Green-function form as \cite{Kleinman1966PR}
\begin{equation}
	\psi_u(\mathbf r)= \frac{\omega_u^2}{c^2}\int_V G_3(\mathbf r,\mathbf r') P_u^{(2)}(\mathbf r')\,d\mathbf r'\ ,
	\tag{C3}
	\label{eqC3}
\end{equation}
where $G_3(\vb{r},\vb{r}')$ is the three-dimensional Green function inside the crystal
\begin{equation}
	G_3(\vb{r},\vb{r}')=\frac{e^{ik_{u}|\vb{r}-\vb{r}'|}}{|\vb{r}-\vb{r}'|}\ .
	\tag{C4}
	\label{eqC4}
\end{equation}

Since both the pump and signal beams are Gaussian, the nonlinear polarization is also Gaussian-like in the transverse plane. Accordingly, the upconverted filed inside the crystal without spatial walk-off and absorption is approximately given by \cite{Kleinman1966PR, Boyd1965PR} 
\begin{equation}
		\psi_u(\mathbf r,t)=C(z)\exp[i(k_u z-w_u t)] \times
		\exp\left( -\frac{x^2+y^2}{w_u^2\Phi_u(z)} \right) . 
\tag{C5}
\label{eqC5}
\end{equation}
Here, $w_{u}$ is the effective beam radius, and $\Phi_u(x,y,z)$ represents the wavefront curvature and other propagation-induced phase terms. The longitudinal buildup factor $C(z)$ is determined by the nonlinear coupling and can be expressed as \cite{Boyd1968JAP,Boyd1965PR}  
\begin{equation}
	C(z)=\kappa\int_0^z G(z')\,e^{i\Delta k_{\mathrm{eff}}z'}\,dz'\ ,
	\label{eqC6} 
	\tag{C6}
\end{equation}
where $\kappa$ is the effective nonlinear coupling coefficient, $G(z)$ denotes the transverse mode-overlap between the pump and signal fields, and $\Delta k_{\text{eff}}$ is the effective phase mismatch including the quasi-phase-matching contribution. In practical nonlinear devices, phase mismatch is compensated through domain-inversion modulation \cite{Chen2018JO}, such that the effective phase-matching condition can be approximated treated as $\Delta k_{\text{eff}}=0$. 

If the waist radii of the signal and pump beams are $w_{s0}$ and $w_{p0}$
, respectively, then the effective waist of the Gaussian nonlinear source is given by
\begin{equation}
w_{u}
=\frac{w_{s0}w_{p0}}
{\sqrt{w_{s0}^2+w_{p0}^2}} \ .
\tag{C7}
\label{eqC7}
\end{equation}
Although Eq. (\ref{eqC5}) and Eq. (\ref{eqC6}) provide a convenient and sufficiently accurate representation for the calculations in the main text, a full explicit evaluation of these expressions is beyond the scope of this work. In the slowly varying envelope approximation and near-phase-matched limit, Eq. (\ref{eqC6}) reduces to an approximately linear dependence on $z$, yielding a familiar $z$-weighted Gaussian-like form for the amplitude.

\end{document}